\documentclass[a4paper,11pt]{article}
\usepackage{pos}
\usepackage{siunitx}
\usepackage{gensymb}
\graphicspath{ {./figures/} }
\DeclareSIUnit\angstrom{\text{\AA}}
\newcommand{\fermi}{{\it Fermi}-LAT }
\usepackage{soul,xcolor}

\title{TeV Detection of the Extreme HSP Blazar RBS 1366 by VERITAS}
\ShortTitle{VERITAS Detection of RBS 1366}

\author*[a,b]{Deivid Ribeiro}
\onbehalf{for the VERITAS Collaboration}

\affiliation[a]{School of Physics \& Astronomy, 116 Church St SE, Minneapolis, USA}
\affiliation[b]{Minnesota Institute for Astrophysics, 116 Church St SE, Minneapolis, USA}

\emailAdd{ribei056@umn.edu}

\setstcolor{red}
\abstract{Extreme high-synchrotron-peak blazars (EHSPs) are postulated as the most efficient and extreme particle accelerators in the universe but remain enigmatic as a possible new class of TeV gamma-ray blazars. Blazars are active galactic nuclei (AGNs) with jets of relativistic particles that generate non-thermal emission pointed along the line-of-sight. Their spectral energy distribution (SED) are characterized by synchrotron and inverse-Compton peaks, indicating acceleration of leptonic and possibly hadronic particle populations in the jet. EHSPs are characterized by a peak synchrotron frequency $> \SI{e17}{Hz}$ with their Compton peak expected to fall in the TeV range. Indeed, the handful of EHSPs detected by Imaging Air Cherenkov Telescopes (IACTs) have presented challenges where some may be a high-frequency extension of the blazar sequence while others peaking around 10 TeV may represent a different class of TeV emitters. Detections of the high-energy and very-high-energy (HE; E > 100 MeV, VHE; E > 100 GeV) components of the Compton peak will play an important role in constraining the acceleration model derived from the SED. We present the discovery of TeV emission from RBS 1366, a candidate EHSP, by the VERITAS observatory. Using HE and VHE data from the \fermi and VERITAS observatories, respectively, we characterize the detection by providing an SED and model fit in the context of other EHSP candidates. Our work confirms the status of RBS 1366 as an EHBL.}

\ConferenceLogo{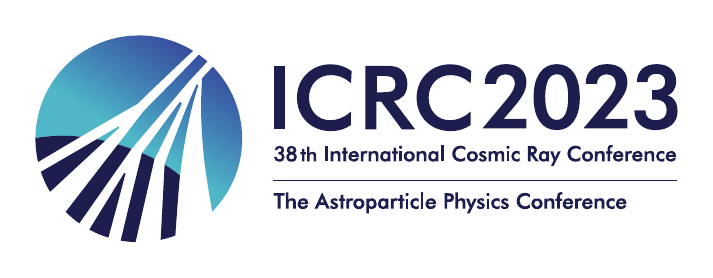}

\FullConference{%
38th International Cosmic Ray Conference (ICRC2023)\\
  26 July - 3 August, 2023\\
  Nagoya, Japan}

\begin{document}
\setstcolor{red}
\maketitle
\section{Introduction}

\sisetup{uncertainty-mode = separate} \sisetup{
output-open-uncertainty = [, output-close-uncertainty = ],
uncertainty-separator
}%

Blazars are a subclass of active galactic nuclei with relativistic jets pointed toward the observer, emitting gamma rays in the  very-high-energy (VHE; E>100 GeV) or "TeV" regime. The underlying mechanisms for the emission of these gamma rays are evident in the observed spectral energy distribution (SED), which is modeled to include the underlying particle populations, and acceleration and cooling mechanisms in the jets. 

It is commonly understood that the low energy and high energy peaks of the SED are produced by synchrotron and inverse Compton processes, respectively \cite{tavecchio1998}. In this synchrotron self-compton model (SSC), a population of electrons emit synchrotron radiation and then the same electron population emit gamma rays through Inverse Compton scattering the synchrotron photons.

At a redshift of $z=0.2365$, RBS 1366 is an AGN in giant elliptical host galaxy with a central black hole mass of $9.31\pm0.32$ M$_{\odot}$ \cite{keenan2021}. RBS 1366 has been a prime candidate for detection and classification as an EHSP \cite{toomey2020,costamante2002,foffano2019}. It was selected for VERITAS observation based on the high synchrotron peak frequency in the Swift XRT X-ray band ($>10^{17}$ Hz), along with detection in the 20 MeV to 300 GeV \fermi band displaying a hard spectrum (and placement in the 3FHL catalog \cite{ajello2017}). RBS 1366 was expected to behave very much like the EHBL 1ES 0229+200 (see Table 1 of \cite{foffano2019}) with a measured a redshift of z = 0.237 based on Ca II, G band, Fe I, Mg I and Na absorption \cite{halpern1986}. In the 1-300 GeV band, Toomey (2020) \cite{toomey2020} found a power law fit with normalization $k = \SI{7.20\pm1.65e-11}{cm^{-2} s^{-1}GeV^{-1}}$ and index $\gamma= \num{1.63\pm0.08}$. 

It is notable that a VERITAS upper limit has been reported \cite{archambault2016, foffano2019}, anticipating that a detection of this source with an accompanying spectrum would enable further EHSP characterization. A differential upper limit of \SI{1.7e-11}{cm^{-1}s^{-1}erg^{-1}} at \SI{327}{GeV} was calculated for this source by VERITAS based on 10 hours of observations in 2016 \cite{archambault2016}.




\section{Methods}\label{sec:methods}

\subsection{Fermi-LAT}\label{subsec:fermilat}
The Large Area Telescope (LAT) on board the {\it Fermi} satellite has operated since 2008 \citep{Atwood2009}. It is sensitive to photons between \SI{{\sim}20}{\MeV} and \SI{{\sim}1}{\TeV} and has an ${\sim}60\degree$ field of view, enabling it to survey the entire sky in about 3 hours. 

RBS 1366 was observed from 2008 to 2023, where the data was fit with a power law as the base spectral model. The \fermi 4FGL catalog defines this source as J1417.9+2543 located at $14^{h}17^{m}58.6^{s}$, \ang{+25;43;26}. 

The publicly available \fermi data were analyzed using with the \texttt{Fermitools} suite of tools provided by the {\it Fermi} Science Support Center (FSSC). Using the \texttt{Fermipy} analysis package \citep{wood2017}\footnote{\url{https://fermipy.readthedocs.io/en/latest/} ; v1.2}, the data were prepared for a binned likelihood analysis in which a spatial spectral model is fit over the energy bins. The data were selected using the SOURCE class of events, which are optimized for point-source analysis, within an angle of $15\degree$ from the analysis target position. A $90\degree$ zenith angle cut was applied to remove any external background events due to the effect of the Earth. The standard background models were applied to the test model, incorporating an isotropic background and a galactic diffuse emission model without any modifications (\textit{$gll\_iem\_v07$} and \textit{$iso\_P8R3\_SOURCE\_V2\_v1$}). The standard 4FGL catalog was then queried for sources within the field of view and their default model parameters \cite{abdollahi2020}. With the improvements to \fermi low-energy sensitivity in PASS8 reconstruction, the energy range was expanded to $100$ MeV - $1$ TeV.

\subsection{VERITAS} \label{subsec:veritas}
The Very Energetic Radiation Imaging Telescope Array System (VERITAS) is an Imaging Atmospheric Cherenkov Telescope (IACT) array consisting of four 12 m telescopes separated by approximately \SI{100}{m}, at the Fred Lawrence Whipple Observatory (FLWO) in southern Arizona, USA \citep{weekes2002, Holder2006}. The observatory is sensitive to photons within the energy range $\backsim$\SI{100}{GeV} to $\backsim$\SI{30}{TeV}, with the ability to detect 1\% of the emission of the Crab Nebula in 25 hours (at $5\sigma$). The instrument has an angular resolution (68\% containment) of $\backsim$0.1\degree\ at \SI{1}{TeV}, an energy resolution of $\backsim$15\% at \SI{1}{TeV}, and 3.5\degree\ field of view.

VERITAS observed RBS 1366 for \SI{56}{hr} between 2008 and 2021, under dark sky conditions. The data in this paper were taken using ``wobble" pointing mode, where the source is offset from the center of the camera by $0.5\degree$. This mode creates space for a radially symmetric "off" region to be used for background estimation in the same field of view, saving time from targeted background observations that contain the same data observing conditions. 

The data were processed with standard VERITAS calibration and reconstruction pipelines, and then cross-checked with a separate analysis chain \citep{cogan2008,Maier2017}. Specifically, we used an Image Template Method (ITM) to improve event angular and energy reconstruction \citep{christiansen2017}, where analysis cuts are determined with a set of a priori data-selection cuts optimized on sources with a soft power law index (from 2.5 to 3).

\section{Results}
\subsection{\fermi}
Using the \fermi data from this work, RBS 1366 (as 4FGL J1417+2543) was detected with a test statistic TS=$353.9$ ($\sim18\sigma$) over an observation period of 14.5 years in the energy range from 100 GeV to 1 TeV. The source was fit to a power law model $dn/dE = (4.13\pm0.76 \times 10^{-15}) (E/10.4 \mathrm{~GeV})^{-1.6\pm0.1}$ MeV$^{-1}$ cm$^{-2}$ s$^{-1}$. The spectrum is shown in the left of Figure \ref{fig:both_specs}.


RBS 1366 was also tested for time variability by constructing a light curve over the entire observing period in 6 month bins, integrating over the same energy range of 100 GeV to 1 TeV. The light curve is binned into periods determined by the Bayesian Blocks algorithm where a false alarm rate of $p0 = 0.0027$ (equivalent to 3 sigma) is used to compute the prior of the number of bins, yielding 3 bins of relative stability, see Figure \ref{fig:vts-lc}\cite{scargle2013}. This is consistent with the 4FGL catalog entry for RBS 1366, where the variability index is measured to be 17.4 (an index above 24.72 over 12 intervals indicates <1\% chance of being a steady source \cite{abdollahi2022}).

\subsection{VERITAS}
\begin{figure}[t]
    \centering
    \includegraphics[height=2in]{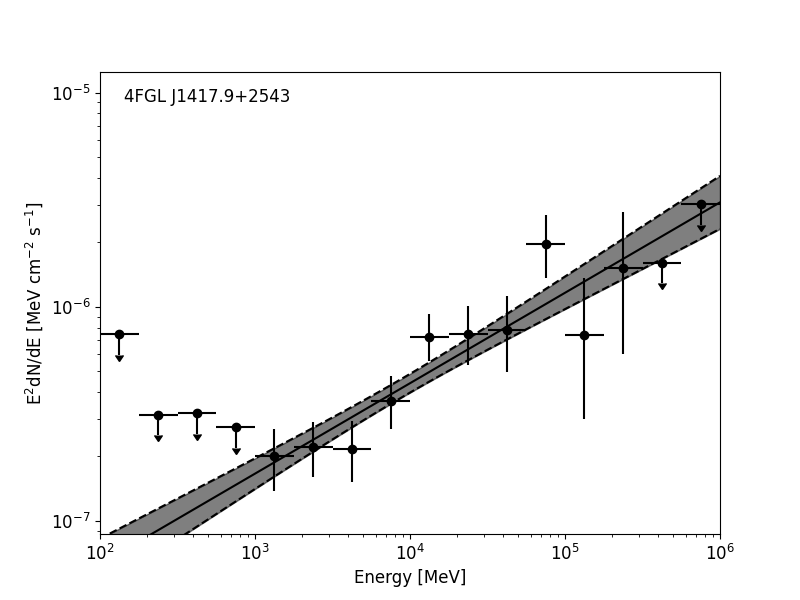}
    \includegraphics[height=2in]{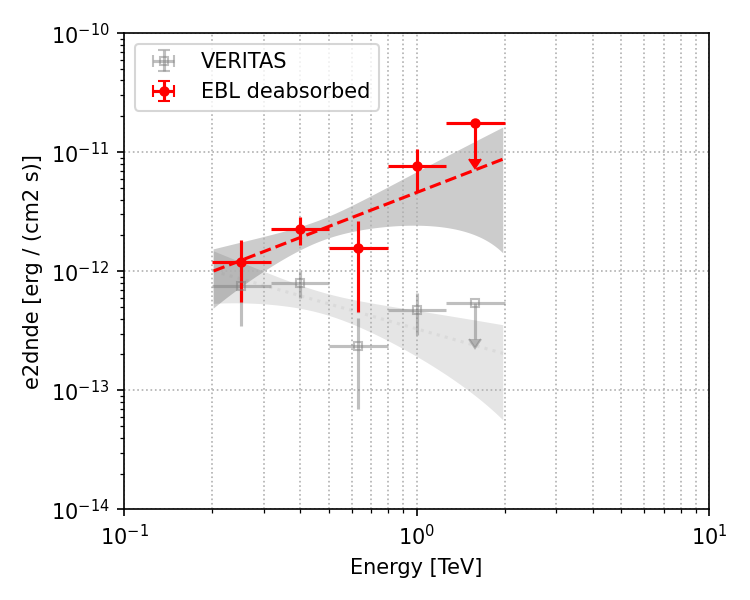}
    \caption{\textit{Left)} Differential spectrum for \fermi source 4FGL J1417+2543, fit using a power law. \textit{Right)} VERITAS spectrum. A TS threshold was set to $1\sigma$. \textit{Grey open square}: observed differential spectrum with power law fit to data. \textit{Red circle}: EBL deabsorbed differential spectrum using the Dominguez model \cite{dominguez2011}, for redshift z=0.237 \cite{halpern1986}. For low significance bins (TS<1), the Rolke upper limit is derived for a power law of index 2.5, and is not included in the overall fit of the larger energy range \cite{rolke2005}. }
    \label{fig:both_specs}
\end{figure}

VERITAS has detected RBS 1366 at a significance of 6.5 sigma or standard deviation, with an exposure of 56.8 hours. A summary of the detection is shown in Table \ref{tab:vts_results}. The observed spectrum was fit to a power law model, shown in Figure \ref{fig:both_specs}.
    
\begin{table}[h]
\centering
\caption{Summary of VERITAS detection. The quality-selected live time, number of gamma-ray-like events in the on- and off-source regions, the normalization for the larger off-source region, the observed excess of gamma-rays and the corresponding statistical significance are shown. For each observation epoch, the integral flux corresponding to the observed excess is given. The flux is reported above the observation threshold of 200 GeV, and is also given in percentage of Crab Nebula flux above the same threshold.}
\begin{tabular}{lcccccccc}
\hline \hline
Epoch     & T    & On   & Off   & Norm   & Excess & Significance & Flux (>200 GeV)           & Crab \\
\hline
          & [hr] &      &       &        &        & [$\sigma$]     & [\SI{e-12}{cm^{-2} s^{-1}}] & \%    \\
2008-2022 & 57   & 3006 & 29180 & 0.0909 & 353    & 6.53         &   $1.7 \pm 0.5$          & 0.5 \\
\hline
\end{tabular}
\label{tab:vts_results}
\end{table}

An effective energy threshold of \SI{\sim200}{GeV} is calculated. The spectrum is fit between \SI{200}{GeV} and \SI{2}{TeV} where the photon count in each bin is above 10 excess counts and the significance is above $1\sigma$, since this source is weakly detected. Upper limit points were calculated to a confidence limit of 99\%. The spectrum is fit with a power law, where upper limits are not included. The power law fit gives ($2.39\times10^{-12} \pm5.2\times10^{-13}) \times (E/400 \mathrm{~GeV})^{(-2.7 \pm 0.5)}$ \unit{TeV^{-1} cm^{-1} s^{-1}}.


\begin{figure}[t]
    \centering
    \includegraphics[width=\textwidth]{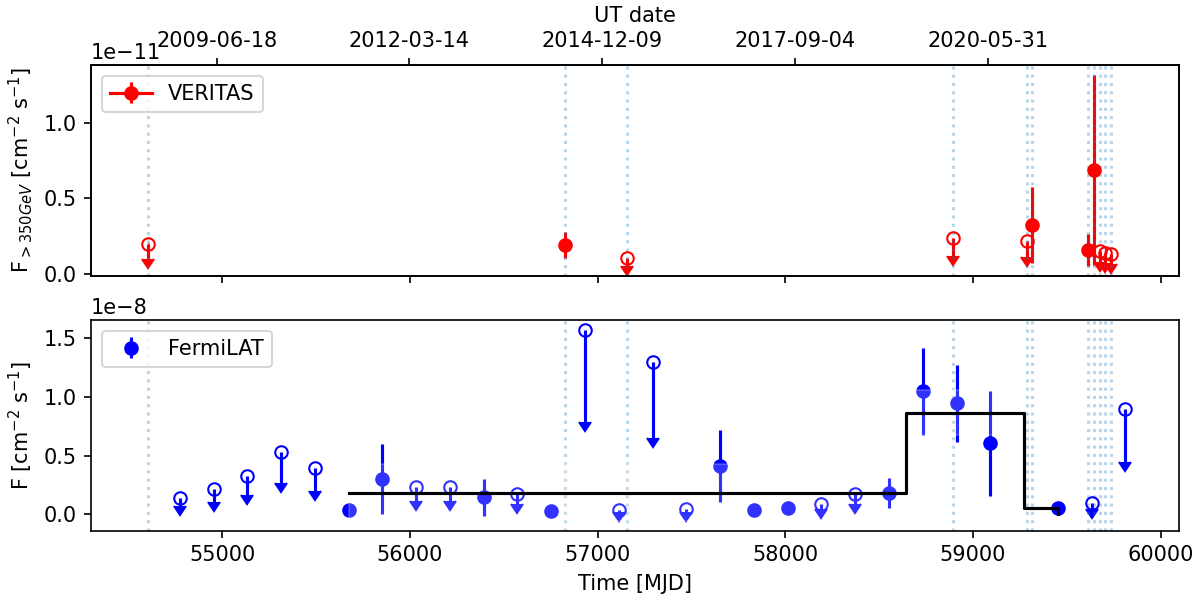}
    \caption{Light curve for VERITAS (red) and \fermi (blue) observations. VERITAS data is binned into 30 day bins, and the flux is calculated above\SI{350}{\GeV}. The \fermi data is binned into 6 month bins and the flux is calculated above \SI{100}{\MeV}. Dotted vertical lines mark the VERITAS observation on the \fermi light curve. A significance threshold of $3\sigma$ was used to determine upper limits, and the Bayesian Block algorithm was used with a false positive probability equivalent to $3\sigma$ to determine bin edges in the \fermi light curve.}
    \label{fig:vts-lc}
\end{figure}

VHE photons are absorbed by the extragalactic background light (EBL) throughout the universe; the flux must be corrected to account for this effect. This absorption is energy and redshift dependent. Deabsorption is applied to the flux using the model of \cite{dominguez2011}. After deabsorbing the spectrum from the EBL, the power law fit is $(7.49\times10^{-12}\pm1.7\times10^{-12} )\times(E/400\mathrm{~GeV})^{(-1.1 \pm 0.6)}$ \unit{TeV^{-1} cm^{-1} s^{-1}}. Although these flux points are harder than the overlapping \fermi data, they are consistent within error. Both observed and EBL deabsorbed spectral points are shown in Figure \ref{fig:both_specs}.

\begin{figure}
    \centering
    \includegraphics{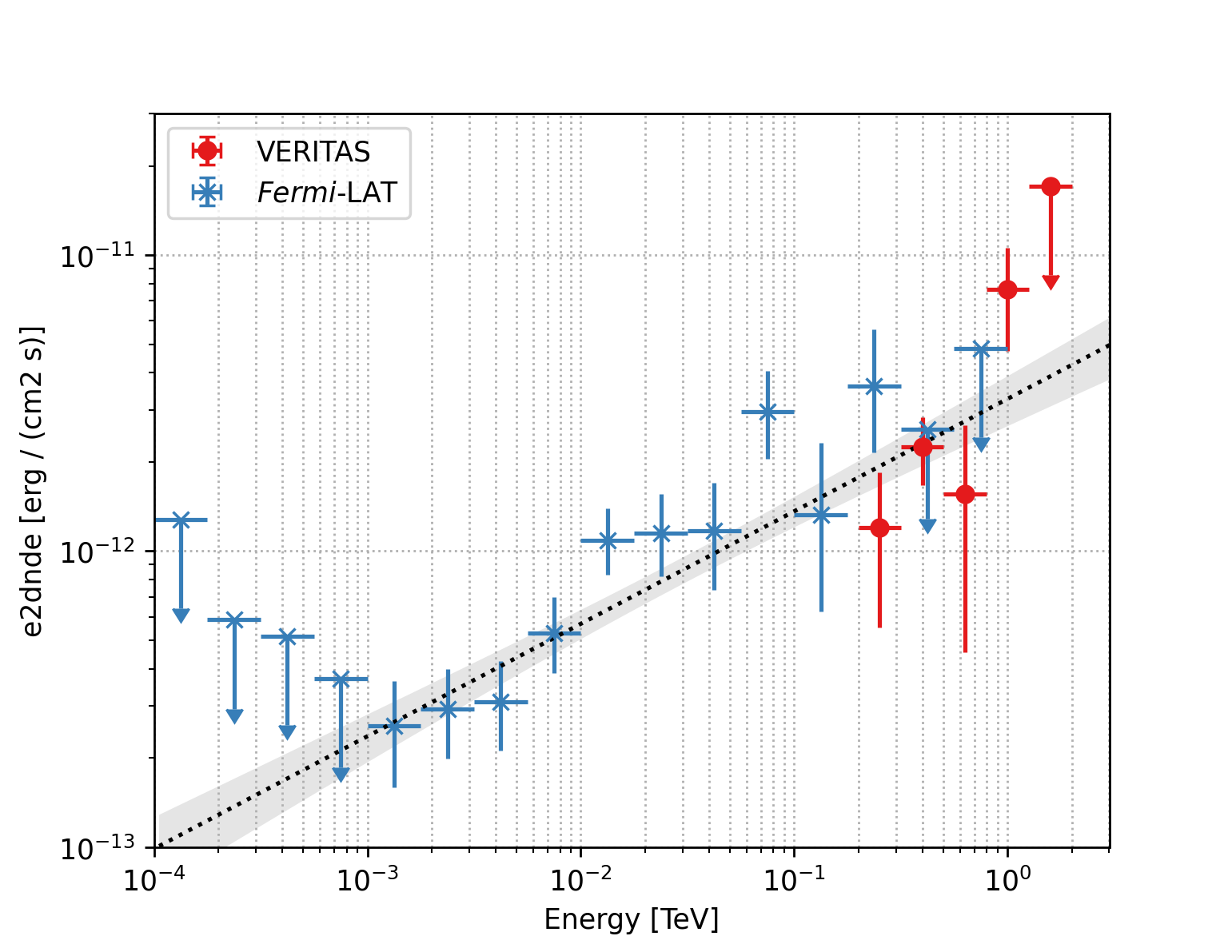}
    \caption{Combined \fermi and VERITAS spectrum with a power law fit (dotted line). The peak of the Compton component is not clear, but is above 0.1 TeV and possibly higher.}
    \label{fig:spec_he_vhe}
\end{figure}

The combined HE and VHE spectral fit is shown in Figure \ref{fig:spec_he_vhe}. Although the VERITAS flux point are hardened by the lowest energy point at 200 GeV, the combined power law fit is consistent within error with a $\chi^2$ statistic of 13.49 (ndof=12). These fit results are summarized in Table \ref{tab:pl_fit}.

\begin{table}[]
\centering
\caption{Power law spectral fit to HE and VHE data.}
\label{tab:pl_fit}
\begin{tabular}{llll}
\hline
Parameter      & Value      & Uncertainty & Unit     \\
Index     & 1.62 & 0.05 &               \\
Amplitude & $9.0$ & $1.4 $ & $10^{-12}$ cm$^{-2}$ s$^{-1}$ TeV$^{-1}$ \\
Reference & 0.4 & -  & TeV           \\\hline
\end{tabular}
\end{table}

In addition to the VERITAS SED, a light curve was also computed over 30 day bins, integrating above 300 GeV. This is shown in Figure \ref{fig:vts-lc} along with the corresponding \fermi light curve. 

\section{Discussion \& Conclusion}
The source exhibited a hard HE+VHE spectrum (index < 2) after correcting for EBL absorption. With a synchrotron peak at log frequency of $17.7\pm0.2$ (\SI{\sim2}{\keV}), a redshift of $0.237$, and the known hardness of the \fermi power law, the evidence already points toward an EHSP classification. The combined HE+VHE fit in this work supports this classification and merits investigation into the subclassification posited by \cite{foffano2019}. The power law index $1.62\pm0.05$ on the combined EBL-deabsorbed HE+VHE spectrum in Figure \ref{fig:spec_he_vhe} gives a VHE gamma-ray slope ($S=2-\gamma$ where $\gamma$ is power law index) of $S=0.38$, which places RBS 1366 similar to other blazars such as 1ES 1101-232, 1ES 0347-121 and 3FGL J0710+5808, which were labeled as Hard-TeV EHBLs in \cite{foffano2019}]. A full multi-wavelength-SED will enable a complete classification in a future study. 

The HE-VHE luminosity of RBS 1366 appears to peak above an energy of $\sim1$ TeV, where it reaches an intrinsic luminosity of $3 \times 10^{44}$ erg/s (assumed to be isotropic). However, since the turnover of the Compton component is not yet evident, we are not able to claim that the Compton component is not dominant over the synchrotron component, which peaks at a log frequency of $17.7\pm0.2$ (\SI{\sim2}{\keV}) with a luminosity of $\sim 10^{45}$ erg/s \cite{foffano2019}. A more sensitive observation from the Cherenkov Telescope Array (CTA), a next generation IACT observatory, in the future may yield a more comprehensive estimation of the Compton peak with improved sensitivity above $1$ TeV \cite{funk2013, thectaconsortium2019}.

A basic estimate of the relativistic boosting in the SSC model fit to available published data is very high, between $\delta \sim50-100$, implying a significantly efficient acceleration that is common to EHSPs. This indicates the emission region is probably in the jet where the emission region is in the range of $10^{17}$ cm and the magnetic field is very weak. While variability has been noted in the optical band \cite{archambault2016}, the lack of variability in the HE-VHE band is expected for the rising end of the Compton peak seen here. 

\begin{figure}
    \centering
    \includegraphics[width=\textwidth]{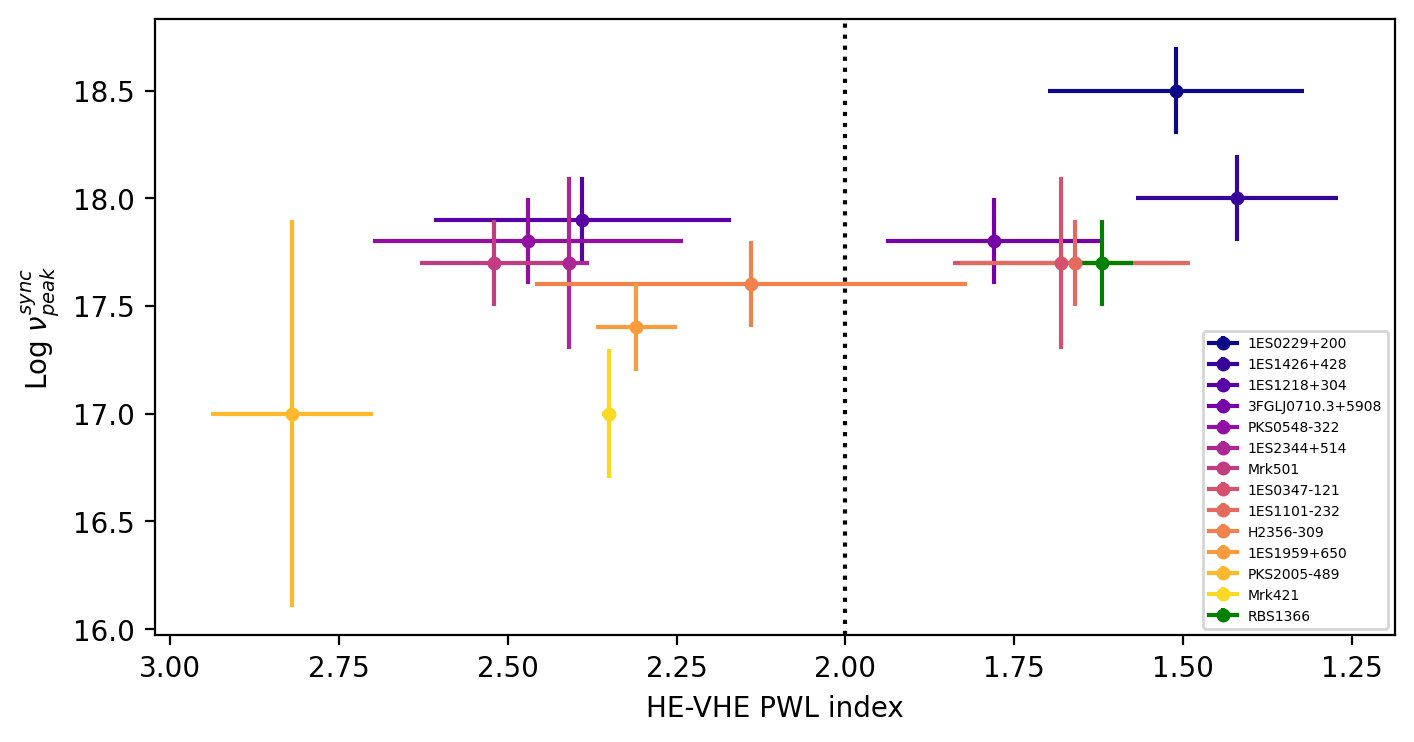}
    \caption{Using a selection of TeV detected EHSPs, the HE-VHE power law index in the range of 0.1 to 10 TeV vs the log frequency of the synchrotron peak. RBS 1366 is shown in green. At these energies, EHSPs diverge into Hard-TeV blazars with high synchrotron peaks (clustered to the right), and HBL-like variable blazars with low synchrotron peak and softer spectrum (clustered to the left). This is reproduced from \cite{foffano2019}. }
    \label{fig:foffano_index_freq}
\end{figure}

The observational circumstances above all support the conclusion that RBS 1366 is an EHSP now detected in the TeV energy range. Figure \ref{fig:foffano_index_freq} is a reproduction of the index versus synchrotron peak frequency plot sampling the other TeV detection EHSPs in \cite{foffano2019}, which cluster into Hard-TeV and HBL-like blazars. Using the results presented in this work, RBS 1366 may be classified as a Hard-TeV EHSP. A complete study using a full multi-wavelength SED to confirm the Hard-TeV EHSP classification is underway. 



\acknowledgments
This work was partially supported by NSF award PHY 2110737. This research is supported by grants from the U.S. Department of Energy Office of Science, the U.S. National Science Foundation and the Smithsonian Institution, by NSERC in Canada, and by the Helmholtz Association in Germany. This research used resources provided by the Open Science Grid, which is supported by the National Science Foundation and the U.S. Department of Energy's Office of Science, and resources of the National Energy Research Scientific Computing Center (NERSC), a U.S. Department of Energy Office of Science User Facility operated under Contract No. DE-AC02-05CH11231. We acknowledge the excellent work of the technical support staff at the Fred Lawrence Whipple Observatory and at the collaborating institutions in the construction and operation of the instrument.

\bibliographystyle{ICRC}
\bibliography{RBS1366_ICRC}

%
%
%

\clearpage

\section*{Full Author List: VERITAS Collaboration}

\scriptsize
\noindent
A.~Acharyya$^{1}$,
C.~B.~Adams$^{2}$,
A.~Archer$^{3}$,
P.~Bangale$^{4}$,
J.~T.~Bartkoske$^{5}$,
P.~Batista$^{6}$,
W.~Benbow$^{7}$,
J.~L.~Christiansen$^{8}$,
A.~J.~Chromey$^{7}$,
A.~Duerr$^{5}$,
M.~Errando$^{9}$,
Q.~Feng$^{7}$,
G.~M.~Foote$^{4}$,
L.~Fortson$^{10}$,
A.~Furniss$^{11, 12}$,
W.~Hanlon$^{7}$,
O.~Hervet$^{12}$,
C.~E.~Hinrichs$^{7,13}$,
J.~Hoang$^{12}$,
J.~Holder$^{4}$,
Z.~Hughes$^{9}$,
T.~B.~Humensky$^{14,15}$,
W.~Jin$^{1}$,
M.~N.~Johnson$^{12}$,
M.~Kertzman$^{3}$,
M.~Kherlakian$^{6}$,
D.~Kieda$^{5}$,
T.~K.~Kleiner$^{6}$,
N.~Korzoun$^{4}$,
S.~Kumar$^{14}$,
M.~J.~Lang$^{16}$,
M.~Lundy$^{17}$,
G.~Maier$^{6}$,
C.~E~McGrath$^{18}$,
M.~J.~Millard$^{19}$,
C.~L.~Mooney$^{4}$,
P.~Moriarty$^{16}$,
R.~Mukherjee$^{20}$,
S.~O'Brien$^{17,21}$,
R.~A.~Ong$^{22}$,
N.~Park$^{23}$,
C.~Poggemann$^{8}$,
M.~Pohl$^{24,6}$,
E.~Pueschel$^{6}$,
J.~Quinn$^{18}$,
P.~L.~Rabinowitz$^{9}$,
K.~Ragan$^{17}$,
P.~T.~Reynolds$^{25}$,
D.~Ribeiro$^{10}$,
E.~Roache$^{7}$,
J.~L.~Ryan$^{22}$,
I.~Sadeh$^{6}$,
L.~Saha$^{7}$,
M.~Santander$^{1}$,
G.~H.~Sembroski$^{26}$,
R.~Shang$^{20}$,
M.~Splettstoesser$^{12}$,
A.~K.~Talluri$^{10}$,
J.~V.~Tucci$^{27}$,
V.~V.~Vassiliev$^{22}$,
A.~Weinstein$^{28}$,
D.~A.~Williams$^{12}$,
S.~L.~Wong$^{17}$,
and
J.~Woo$^{29}$\\
\\
\noindent
$^{1}${Department of Physics and Astronomy, University of Alabama, Tuscaloosa, AL 35487, USA}

\noindent
$^{2}${Physics Department, Columbia University, New York, NY 10027, USA}

\noindent
$^{3}${Department of Physics and Astronomy, DePauw University, Greencastle, IN 46135-0037, USA}

\noindent
$^{4}${Department of Physics and Astronomy and the Bartol Research Institute, University of Delaware, Newark, DE 19716, USA}

\noindent
$^{5}${Department of Physics and Astronomy, University of Utah, Salt Lake City, UT 84112, USA}

\noindent
$^{6}${DESY, Platanenallee 6, 15738 Zeuthen, Germany}

\noindent
$^{7}${Center for Astrophysics $|$ Harvard \& Smithsonian, Cambridge, MA 02138, USA}

\noindent
$^{8}${Physics Department, California Polytechnic State University, San Luis Obispo, CA 94307, USA}

\noindent
$^{9}${Department of Physics, Washington University, St. Louis, MO 63130, USA}

\noindent
$^{10}${School of Physics and Astronomy, University of Minnesota, Minneapolis, MN 55455, USA}

\noindent
$^{11}${Department of Physics, California State University - East Bay, Hayward, CA 94542, USA}

\noindent
$^{12}${Santa Cruz Institute for Particle Physics and Department of Physics, University of California, Santa Cruz, CA 95064, USA}

\noindent
$^{13}${Department of Physics and Astronomy, Dartmouth College, 6127 Wilder Laboratory, Hanover, NH 03755 USA}

\noindent
$^{14}${Department of Physics, University of Maryland, College Park, MD, USA }

\noindent
$^{15}${NASA GSFC, Greenbelt, MD 20771, USA}

\noindent
$^{16}${School of Natural Sciences, University of Galway, University Road, Galway, H91 TK33, Ireland}

\noindent
$^{17}${Physics Department, McGill University, Montreal, QC H3A 2T8, Canada}

\noindent
$^{18}${School of Physics, University College Dublin, Belfield, Dublin 4, Ireland}

\noindent
$^{19}${Department of Physics and Astronomy, University of Iowa, Van Allen Hall, Iowa City, IA 52242, USA}

\noindent
$^{20}${Department of Physics and Astronomy, Barnard College, Columbia University, NY 10027, USA}

\noindent
$^{21}${ Arthur B. McDonald Canadian Astroparticle Physics Research Institute, 64 Bader Lane, Queen's University, Kingston, ON Canada, K7L 3N6}

\noindent
$^{22}${Department of Physics and Astronomy, University of California, Los Angeles, CA 90095, USA}

\noindent
$^{23}${Department of Physics, Engineering Physics and Astronomy, Queen's University, Kingston, ON K7L 3N6, Canada}

\noindent
$^{24}${Institute of Physics and Astronomy, University of Potsdam, 14476 Potsdam-Golm, Germany}

\noindent
$^{25}${Department of Physical Sciences, Munster Technological University, Bishopstown, Cork, T12 P928, Ireland}

\noindent
$^{26}${Department of Physics and Astronomy, Purdue University, West Lafayette, IN 47907, USA}

\noindent
$^{27}${Department of Physics, Indiana University-Purdue University Indianapolis, Indianapolis, IN 46202, USA}

\noindent
$^{28}${Department of Physics and Astronomy, Iowa State University, Ames, IA 50011, USA}

\noindent
$^{29}${Columbia Astrophysics Laboratory, Columbia University, New York, NY 10027, USA}

\end{document}